\documentclass[referee,sn-aps]{sn-jnl}

\usepackage{graphicx}%
\usepackage{multirow}%
\usepackage{amsmath,amssymb,amsfonts}%
\usepackage{amsthm}%
\usepackage{mathrsfs}%
\usepackage[title]{appendix}%
\usepackage{xcolor}%
\usepackage{textcomp}%
\usepackage{manyfoot}%
\usepackage{booktabs}%
\usepackage{listings}%
\newcommand\beq{\begin{eqnarray}}
\newcommand\eeq{\end{eqnarray}}
\newcommand\bq{\begin{equation}}
\newcommand\eq{\end{equation}}

\begin{document}

\title[Article Title]{  Recoil electron polarization-dependent T-odd correlations   in  neutrino   elastic scattering on polarized electron target}


\author[1]{\fnm{Arkadiusz} \sur{  B\l{}aut}}\email{arkadiusz.blaut@uwr.edu.pl}
\equalcont{These authors contributed equally to this work.}

\author*[1]{\fnm{{Wies\l{}aw} \sur Sobk\'ow}}\email{wieslaw.sobkow@@uwr.edu.pl}
\equalcont{These authors contributed equally to this work.}


\affil*[1]{\orgdiv{Institute of Theoretical Physics}, \orgname{University of Wroc\l{}aw}, \orgaddress{\street{Pl. M. Born 9}, \city{Wroc\l{}aw}, \postcode{50204}, \country{Poland}}}




\abstract{Possible symmetry breaking tests with respect to the time inversion in the elastic scattering of neutrinos on polarized electrons are considered, assuming that  the incoming neutrino beam is either longitudinally or transversely polarized, and  both momentum and polarization of recoil electrons are observed. 
 In the process, in addition to the standard interaction, the exotic scalar, pseudoscalar and tensor interactions can participate. Due to the nonstandard interactions, different types of triple  T-odd products in the cross section may appear. We consider several experiments in which mixed   products  built of the recoil electron polarization and two other vector quantities (incoming neutrino momentum, its polarization,  polarization of the electron target and outgoing  electron momentum) can be identified. 
 Observations of the T-odd correlations may indicate noninvariance under time reversal as well as the presence of exotic interactions. A complete analysis of the problem requires a precise determination of the possible contributions from the  interactions mimicking the genuine violation of  time reversal symmetry, e.g. final state interactions.}




\maketitle

\section{Introduction}\label{sec1}

 As is known, in any CPT-invariant theory CP symmetry violation (CPV) implies time reversal symmetry violation (TRV) and vice versa. According to the standard model (SM) \cite{SM,SM1,SM2,SM3,SM4} the CPV is described by a single phase of the Cabibbo-Kobayashi-Mas\-ka\-wa quark-mixing matrix (CKM) \cite{Kobayashi}.  Breaking of this symmetry was confirmed in the decays of neutral kaons and B mesons \cite{CP,CP1,CP2}. The first direct observation of the TRV was carried out by the BaBar Collaboration at SLAC \cite{BaBar}, and the first test of T, CP and CPT symmetries performed in transitions of neutral kaons were obtained by the KLOE-2 Collaboration \cite{KLOE2}. 

A natural question arises whether there are sources of CP-symmetry violation other than CKM phase. This issue may turn out to be of primary interest in the early time cosmology, because the measured CKM phase does not explain the observed matter-antimatter asymmetry of the universe \cite{barion}. There are various observables and experimental strategies proposed to detect new sources of CPV and TRV. The historically first group of tests concerns the measurements of T-odd angular correlations  in the decay of nuclei (neutrons) and kaons. One can recall here the project of the BRAND experiment \cite{BRAND} which will measure all correlation coefficients related to the transverse polarization of recoil electrons coming from the polarized neutron decay. The theoretical basis for nuclei decay experiments was first put forward by Jackson et.al. \cite{Jackson} and Ebel \cite{Ebel}. However these tests could be regarded as ambiguous, due to 
possible T-even final state interactions mimicking TRV effects \cite{Jackson1}.
To overcome this obstacle, experiments capable of testing the TRV in an unambiguous way were proposed, such as measurements of the electric dipole moments \cite{NDM,NDM1,NDM2,NDM3,NDM4}, or direct measurements of the reciprocity relation breaking \cite{reci}.
The later includes e.g. neutrino oscillation experiments, which open the possibility of CPV and TRV in the leptonic sector \cite{Geer, Geer1}. 
 

In this work we indicate advantages of neutrino elastic scattering experiments on  polarized target in search of TRV with exotic scalar, tensor and pseudoscalar interactions by the measurements of T-odd correlations, when the polarization of recoil electrons is observed. 
We stress that a complete analysis of the problem requires a precise determination of the possible contributions from the  final state interactions to extract the information about the genuine violation of  time reversal symmetry. In this  study the potential contributions from the  final state interactions are neglected. The present paper can be treated as an extension of the research started in previous work \cite{SBPET}. 
Recall that   according to SM only the vector V and axial A couplings of left chiral neutrinos can take part in elastic scattering of  neutrinos on  electrons and they are real due to the hermiticity conditions of interaction lagrangian. This means that there is no room for T-odd triple angular correlations in the elastic scattering cross section.  The situation clearly changes when complex exotic scalar, tensor and pseudoscalar couplings are introduced and the  scattering target is polarized; measurement capabilities increase further when the recoil electron polarization is measured. In this case the time reversal symmetry breaking triple (mixed) products may appear. 
The present analysis and the construction of new T-odd observables in the context of nonstandard interactions are motivated by the work of Jackson et.al. \cite{Jackson}.

The idea of using a polarized target have already found many interesting applications, e.g. in probing the neutrino magnetic moments \cite{PET,PET4} or the flavor composition of neutrino beam \cite{PET3}, testing TRV \cite{PET2,SBPET} and neutrino nature \cite{PETNN,Barranco,Barranco1},  searches of axions, analysis of spin-spin interaction in gravitation \cite{PET1,PET5,PET6,PET7}, and in the dark matter detection \cite{DM1,DM2,DM3,PDM0}. Similarly, the experiments with polarized target may  be useful in testing the predictions of  many nonstandard  models:
the left-right symmetric models \cite{Pati,Pati1,Pati2,Pati3,Pati4}, composite models \cite{Jodidio,CM,CM1}, models with extra dimensions  \cite{Extra} and the unparticle  models  \cite{unparticle,unparticle1,unparticle2}. 
An added benefit of using a polarized target is the ability to precisely measure the background level, since the contributions to the cross section  can be controlled by changing the direction of  magnetic field \cite{Misiaszek}. In the context of the present paper we take note of preliminary tests of the availability of polarized target  for the neutrino measurements \cite{INFN}. Moreover, methods of producing polarized gases such as helium, argon, xenon  have long been known,  but their application to neutrino experiments can be a huge challenge  \cite{Gass1,Gass2}. 

In the paper we discuss several experimental scenarios for detecting angular correlations  violating time reversal. 
In Sec. \ref{sec2}  we introduce experimental setting and give theoretical description of the considered process.
In Sec. \ref{sec3} we analyze elastic neutrino scattering on polarized target in the case of longitudinally polarized neutrinos.
In Sec. \ref{sec4} we consider the scattering with neutrinos having nonzero transversal spin  component.  Sec. \ref{sec5} gives summary of the results.
Appendixes contain the results for the cross sections of the analyzed scenarios. 

The studies are carried out for the flavor neutrino eigenstates in  the relativistic Dirac neutrino limit.

\section{Detection process --- elastic scattering of neutrinos on polarized electrons}\label{sec2}
In this section we set basic assumptions regarding analyzed scenarios and performed calculations. Our  considerations are based on the process of elastic scattering of neutrinos on  polarized electrons. 
In the experiment with a polarized target we have essentially four vector quantities; the incoming neutrino momentum, the target polarization, the momentum and  polarization of the scattered electron. Each of these quantities changes the sign under   time reversal, so the mixed product of any three of them is invariant with respect to rotations, but noninvariant under time reversal.

It is assumed that the incoming Dirac low energy neutrino beam (produced by  a hypothetical high-intensity emitter located close to the detector) is a superposition of left chiral states with right chiral ones. This means that nonvanishing transversal  components of neutrino spin polarization may appear. The issue of neutrino source selection is beyond the purpose of the work and will be crucial in the conditions of a specific experiment.  
For illustrative purposes, we can indicate the process of the muon capture by proton in which the transversal components of $\nu$ polarization (both time reversal symmetry invariant  and time reversal symmetry breaking) can  appear \cite{CMS2003}. These neutrino observables  consist only of the interference terms  between the standard vector, axial couplings of left chiral neutrinos and   exotic scalar, tensor, pseudoscalar  couplings of right chiral neutrinos and do not vanish in the relativistic $\nu$ limit. 
 \par
\begin{figure*}[h]
	\centering
	\includegraphics[width=1.0\textwidth]{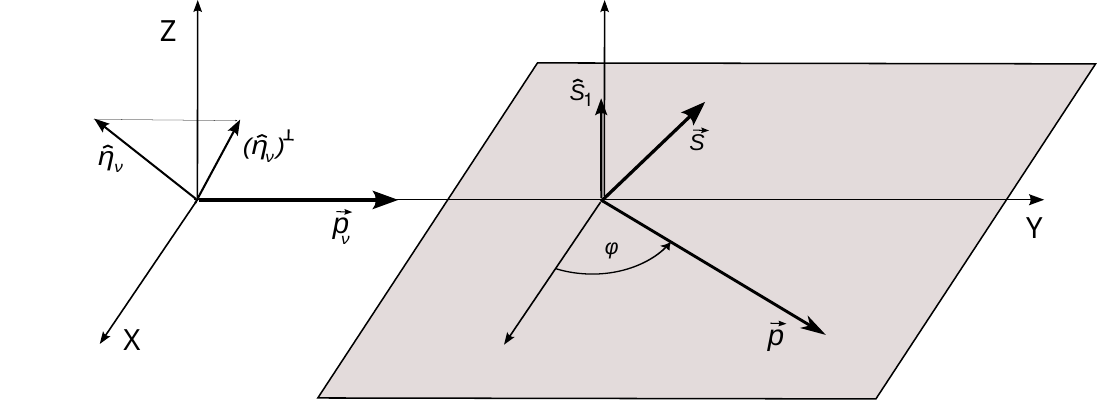}
	\caption{The neutrino scattering on the polarized electron: $\hat{\eta }_{\nu }$ is the unit polarization vector of incoming neutrino in its rest frame. 
	$\hat{\eta}_{\nu}^{\perp}\equiv(\hat{\eta}_{\nu})^{\perp}$ is the transversal component of $\hat{\eta }_{\nu }$;  $\hat{s}_1$ is the unit electron polarization vector of the target, $\vec{s}$ is the recoil electron polarization vector, $\vec{p}$ is the recoil electron momentum, $\varphi$ is the azimuthal angle of the outgoing electron momentum.} \label{fig1}
\end{figure*}

 Fig. \ref{fig1} describes all quantities used in the analysis, and illustrates the relative orientation of the relevant vectors.
The left chiral neutrinos interact by the standard vector-axial interaction (VA) and tiny admixture of nonstandard scalar $\rm S_L$, tensor 
$\rm T_L$ and pseudoscalar $\rm P_L$ interactions. The right chiral neutrinos are only  detected by the exotic scalar $\rm S_R$, tensor 
$\rm T_R$, pseudoscalar $\rm P_R$ interactions. 
The amplitude for the  neutrino-electron scattering at low energies has the form: 
\beq \label{ampD} M_{\nu e^{-}}
&=&
\frac{G_{F}}{\sqrt{2}}\{(\overline{u}_{e'}\gamma^{\alpha}(c_{\rm V}
- c_{\rm A}\gamma_{5})u_{e}) (\overline{u}_{\nu'}
\gamma_{\alpha}(1 - \gamma_{5})u_{\nu})\nonumber\\ 
&  & \mbox{} +
c_{\rm S_{R}}(\overline{u}_{e'}u_{e})(\overline{u}_{\nu'}
(1 + \gamma_{5})u_{\nu}) \nonumber\\
&  & \mbox{} +
c_{\rm P_{R}}(\overline{u}_{e'}\gamma_{5}u_{e})(\overline{u}_{\nu'}
\gamma_{5}(1 + \gamma_{5})u_{\nu}) \nonumber\\
&& \mbox{} +
\frac{1}{2}c_{\rm T_{R}}(\overline{u}_{e'}\sigma^{\alpha \beta}u_{e})(\overline{u}_{\nu'}
\sigma_{\alpha \beta}(1 + \gamma_{5})u_{\nu})\nonumber\\
&&\mbox{} +
c_{\rm S_{\rm L}}(\overline{u}_{e'}u_{e})(\overline{u}_{\nu'}
(1 - \gamma_{5})u_{\nu}) \nonumber\\
&&\mbox{} + c_{\rm P_{L}}(\overline{u}_{e'}\gamma_{5}u_{e})(\overline{u}_{\nu'}
\gamma_{5}(1 - \gamma_{5})u_{\nu}) \nonumber\\
&& \mbox{} +
\frac{1}{2}c_{\rm T_{L}}(\overline{u}_{e'}\sigma^{\alpha \beta}u_{e})(\overline{u}_{\nu'}
\sigma_{\alpha \beta}(1 - \gamma_{5})u_{\nu})
\},  
\eeq
where $G_{F}$ is the Fermi constant \cite{Data}. 
The coupling constants for the nonstandard interactions are
denoted with the subscripts L and R as  $c_{\rm S_{R, L}}$, $c_{\rm P_{R, L}}$,  $c_{\rm T_{R, L}}$ respectively to the incoming $\nu$ 
of left- and right-handed chirality.  These constants are complex numbers denoted as 
$c_{\rm S_R} = |c_{\rm S_R}|e^{i\,\theta_{\rm S,R}}$, $c_{\rm S_L} = |c_{\rm S_L}|e^{i\,\theta_{\rm S,L}}$, etc.  
Hermiticity of the interaction lagrangian generates the relations between the exotic couplings:   
$c_{\rm (S, T, P)_ L}=c_{\rm (S, T, P)_R}^{*}$. 
\par The main goal is to calculate the differential  cross section for the neutrino-electron scattering in the presence of exotic interactions, assuming that the incoming neutrino beam is polarized, the electron target is oriented, and that the polarization of the scattered electrons is measured. This means that the covariant projection operators for polarized particles \cite{Michel} are used. 

\section{Time reversal violation ---  case of longitudinal neutrino polarization }\label{sec3}


In this section we consider $\rm VA, S_L, T_L$ interactions and analyze the effects of  TRV assuming that the incoming beam of longitudinally polarized neutrinos scatters on the polarized electron target and the recoil electron momentum and its polarization are measured.
As a result triple products can appear multiplied by the imaginary part of the products between the exotic scalar and tensor couplings or between the  tensor and pseudoscalar couplings. 
We  discuss various experimental settings allowing unambiguous detection of the signals generated by triple angular correlations.  
To this end we assume that the electron target is polarized perpendicularly to the direction of the incoming neutrino beam ($\hat{p}_\nu \cdot \hat{s}_1=0$).
To extract terms breaking the time reversal symmetry we further assume that the scattered electrons are recorded only in the plane perpendicular to the polarization of the electron target  ($\hat{p} \cdot \hat{s}_1=0$), and events are  collected for the two opposite  directions of the  target polarization. The difference of cross sections taken for $\pm\hat{s}_1$ denoted by $\Omega$, eliminates many terms that do not depend on  $\hat{s}_1$ vector; 
the relevant formulas can be found in the appendix \ref{app1}, Eqs. (\ref{a1VA})-(\ref{a1ST}). The cross section $\Omega$ consisting of the quadratic terms 
$|\rm M_{VA}|^2$, $|\rm M_{S_L}|^2$, $|\rm M_{T_L}|^2$ and the interference term $2\Re(\rm M_{S_L} M^*_{T_L})$, has the structure 
${\rm A}\,\hat{s}_ 1\cdot \vec{s} + {\rm B}\,\hat{p}_ {\nu}\cdot (\hat{s}_1\times \vec{s}) + {\rm C}\,\hat{p}\cdot (\hat{s}_1 \times \vec{s})$.
There are three possible ways to eliminate contributions from all time reversal  invariant terms. 

In the first method one assumes the condition $\hat{s}_1\cdot\vec{s}=0$ which amounts to the detection of the outgoing electron polarization in the scattering plane.
This removes all supposedly dominating quadratic terms, and leaves us solely with the two triple products
${\rm B}\,\hat{p}_ {\nu}\cdot (\hat{s}_1\times \vec{s}) + {\rm C}\,\hat{p}\cdot (\hat{s}_1 \times \vec{s})$.
With an additional restriction that the experiment registers only the orthogonal polarization component of the scattered electron (i.e. 
$\vec{s}=\pm \hat{s}_1\times \hat{p}$), one obtains for the interference
\begin{eqnarray}
2\Re(\rm M_{S_L} M^*_{T_L}) & = & \pm
16 |c_{\rm S_R}||c_{\rm T_R}| m E_{\nu}^{3/2}\sqrt{y} \sin{(\theta_{\rm T,R}-\theta_{\rm S,R})}(2m+E_{\nu}y) \nonumber\\
&&
\left(\sqrt{2m+E_{\nu}y} - \sqrt{E_{\nu}y}\sin{\varphi}\right)
\end{eqnarray}
(the relevant symbols are defined in the appendix \ref{app1}). In the above expression $\hat{p}\cdot (\hat{s}_1 \times \vec{s})$ produces the constant term whereas the azimuthal angle dependece comes from the mixed product 
$\hat{p}_ {\nu}\cdot (\hat{s}_1\times \vec{s})$.

The second method employs the observable $\Omega(\hat{p})-\Omega(-\hat{p})$ in the experiment counting events coming from two opposite momentum  directions of the recoil electrons. As in the previous case the quadratic contributions are annihilated, and only $\hat{p}$-dependent mixed product
$\hat{p}\cdot (\hat{s}_1 \times \vec{s})$ survives. 

The third approach relies on the restriction of the  polarization  measurement of the recoiled electron to its parallel component. In this case $\hat{s}_1\cdot\vec{s}=0$
and $\hat{p}\cdot (\hat{s}_1 \times \vec{s})=0$, thus one is left only with the $\hat{p}_{\nu}$-dependent triple product $\hat{p}_ {\nu}\cdot (\hat{s}_1\times \vec{s})$.

We conclude with the remark valid for each of the above method that any nonzero signal indicates the presence of the nonstandard time reversal violating $\rm S_L$, $\rm T_L$ interactions.
It can be also seen that the contributions responsible for the time reversal violation survive, even if the cross section is integrated 
over the recoil electron energy. In consequence, this may improve the detectability of the expected effect. 

\section{Time reversal violation --- scenario with  transversal neutrino polarization }\label{sec4}
In this section we consider the possibility of time reversal violation in the presence of  $\rm VA, S_L, S_R$ interactions when the incoming neutrino beam has the nonzero transversal polarization components and scatters on the polarized electron target. It is assumed that both the recoil electron momentum and its polarization are measured. In this case, mixed products containing the transverse neutrino polarization vector appear, which are linear in exotic couplings. 
To isolate the time reversal violation contributions we impose the following conditions defining the experimental setup,
$\hat{p}\cdot\hat{s}_1=0$, $\hat{p}\cdot\hat{s}=0$, $\hat{s}_1\cdot\hat{s}=0$ (i.e. the three vectors, target polarization vector, outgoing electron polarization and outgoing electron momentum, form an orthonormal basis). 

In the first step in analogy with \cite{Jackson,Jackson1,Ebel} we consider probabilities  of emission of electrons having spins in two opposite directions (definition of the relevant cross section denoted by $\Delta$ is given in the appendix \ref{app2}). This observable eliminates exotic squared amplitude $|M_{\rm S_R}|^2$ and the interference between  $S_L$ and $S_R$ couplings leaving the standard $|M_{\rm VA}|^2$ proportional to $\hat{p}_{\nu}\cdot \vec{s}$. The resulting interference term in the cross section includes linear combination of all possible triple products,
$ \hat{s}_1 \cdot(\hat{\eta}_{\nu}^{\perp}\times \vec{s})$,
$\hat{p}_{\nu} \cdot(\hat{\eta}_{\nu}^{\perp}\times \vec{s})$,
$\hat{p}\cdot(\hat{\eta}_{\nu}^{\perp}\times \vec{s})$,
$\hat{p}_{\nu} \cdot(\hat{s}_1\times \vec{s})$,
$\hat{p} \cdot(\hat{s}_1\times \vec{s})$; explicit  formulas can be found in the appendix \ref{app2}. 
For unambiguous detection of the exotic $\rm VA-S_R$ interference one should be able to eliminate a dominating strong contribution from the standard squared amplitude $|M_{\rm VA}|^2$. 

In the second step one attempts to get rid of $|M_{\rm VA}|^2$ contribution and separate different time reversal breaking contributions coming from specific triple products. Eq. (\ref{vs3}) shows that a possible way is to assume $\hat{p}_{\nu}\cdot \vec{s}=0$, but this restriction is too severe (it holds when the scattered electrons are parallel to the incoming neutrinos) and renders this method very inefficient.
In the second approach one further specialize the experimental setting taking a difference of cross sections for two opposite directions of target polarization $\pm\hat{s}_1$ (definition of the relevant cross section denoted by $\tilde{\Delta}$ is given in the appendix \ref{app2}). 
As a result $|M_{\rm VA}|^2$ vanishes and one is left solely with a linear combination of 
$\hat{s}_1$-dependent  triple products, $ \hat{s}_1 \cdot(\hat{\eta}_{\nu}^{\perp}\times \vec{s})$,
$\hat{p}_{\nu} \cdot(\hat{s}_1\times \vec{s})$,
$\hat{p} \cdot(\hat{s}_1\times \vec{s})$ (see Eq. (\ref{vs3s1})). One then concludes that any nonzero result of the experiment
confirms the  presence of the T-odd correlations (and exotic TRV $S_{\rm R}$ interactions, under the assumption that the 
final state interactions are negligible).

With the hypothetical assumption of the source producing neutrinos with a fixed (but unknown) orientation of the transversal polarization, one can determine the value of the product $|\hat{\eta}_{\nu}^{\perp}|\Im{(c_{S_R})}$. This can be achieved only when the orientation of
$\hat{\eta}_{\nu}^{\perp}$ polarization vector is known. To identify the direction of $\hat{\eta}_{\nu}^{\perp}$ one repeats the experiment with $\hat{s}_1$ rotated in the plane orthogonal to $\hat{p}_{\nu}$ keeping conditions $\hat{p}\cdot\hat{s}_1=0$, $\hat{p}\cdot\hat{s}=0$, $\hat{s}_1\cdot\hat{s}=0$, unchanged.  
It is now straightforward to recognize a direction $\hat{s}_1$ parallel to $\hat{\eta}_{\nu}^{\perp}$: indeed, in this case both $ \hat{s}_1 \cdot(\hat{\eta}_{\nu}^{\perp}\times \vec{s})$ and $\hat{p}\cdot\hat{\eta}_{\nu}^{\perp}$ vanish, so the net result is zero.
Denoting now by $\varphi_{\nu}$ an angle between $\hat{s}_1$ and $\hat{\eta}_{\nu}^{\perp}$ (oriented from $\rm Z$ to $\rm X$ axis, Fig. \ref{fig1}) one obtains the following formula for the interference:
\begin{eqnarray}
2\Re(\rm M_{VA} M^*_{S_R}) & = &
-8 |\hat{\eta}_{\nu}^{\perp}|\Im{(c_{S_R})} m \sqrt{y}\cos{\varphi}\sin{\varphi_{\nu}} 
\Bigl( \sin{\varphi}\bigl( E_{\nu}\left(2m+E_{\nu}y)\right)^{3/2} \nonumber \\
&& -E_{\nu}\sqrt{y}\left(m+E_{\nu}\right)\left(2m+E_{\nu}y\right) \Bigr).
\end{eqnarray}

In summary, knowing the orientation of the neutrino polarization vector the only quantity to fit the data is $|\hat{\eta}_{\nu}^{\perp}|\Im{(c_{S_R})}$.
Similarly, as in the scenario considered in Sec. \ref{sec3}, the cross section can be integrated over the outgoing electron energy, 
potentially increasing the number of events. 

\section{Conclusions}\label{sec5}

The paper investigated several scenarios for the time reversal violation in the elastic  scattering of Dirac  neutrinos on polarized electrons, with the incoming neutrino beam being either longitudinally or transversely polarized, under the assumption that the momentum and polarization of the scattered electrons are observed. 

In the case of longitudinal neutrino  polarization, it is possible to unambiguously isolate two angular correlations 
that break the symmetry of time reversal. First T-odd term, $\hat{p}_{\nu}\cdot(\hat{s}_1\times\vec{s})$, is composed of the incoming neutrino momentum, target polarization and polarization of the scattered electron. Second T-odd product, $\hat{p}\cdot(\hat{s}_1\times\vec{s})$, consists of recoil electron momentum, its polarization, and polarization of the target.  In the case analyzed in the paper both mixed products are the effects of the interference between the scalar and tensor interactions necessarily involving imaginary part of the coupling constants, $\Im{(c_{S_R})}$ or
$\Im{(c_{T_R})}$. 

For the transverse neutrino  polarization one can, again, clearly identify the contribution from the time reversal symmetry violation triple products. The transversal neutrino polarization enters three mixed products, 
$ \hat{s}_1 \cdot(\hat{\eta}_{\nu}^{\perp}\times \vec{s})$,
$\hat{p}_{\nu} \cdot(\hat{\eta}_{\nu}^{\perp}\times \vec{s})$,
$\hat{p}\cdot(\hat{\eta}_{\nu}^{\perp}\times \vec{s})$, and appears in various double products multiplying two previously mentioned time reversal symmetry breaking terms.
All new effects are linear in the exotic scalar  $S_R$  interaction of right chiral neutrinos (a similar regularity shows up for tensor $T_R$ or pseudoscalar $P_R$ interactions). As in the previous scenario the T-odd terms must necessarily involve the imaginary part of $c_{S_R}$ (or $c_{T_R}$ and $c_{P_R}$).

In both cases the structure of the T-odd correlations would potentially signal breaking of the time reversal symmetry. 
However before drawing the conclusion one should take into account the possibility that T-odd products may be induced by final state interactions 
in a time reversal invariant theory. An analysis of those effects is beyond the scope of the present paper.

The high-precision tests considered in this work would be a major challenge for experimental groups, 
because strong sources of low-energy (monoenergetic) neutrinos, large polarized targets and detectors sensitive to the directionality of scattered electrons are demanded. Besides, measurements of the recoil electron polarization  would also be a difficult task, although the detection methods by Mott scattering have been known for a long time and used in  beta decay experiments. 
Potential measurement of effects associated with transverse neutrino polarization would require further research on the selection of a suitable neutrino source. In particular it is crucial to explain the role of exotic interactions in the production of neutrino beam with nonzero  transverse polarization.

\backmatter




\begin{appendices}
\section{ Scenario with $\rm VA,  S_{L,R}, T_{L,R}$ interactions for longitudinally polarized neutrino}
\label{app1}
The laboratory differential cross section for $\hat{p}_{\nu }\cdot \hat{\eta }_{\nu }=-1$, $\hat{p}_\nu\cdot \hat{s}_1=0$ and $\hat{p}\cdot\hat{s}_1=0$ is defined as
\begin{equation}
\frac{d^2 \sigma}{d\varphi d \,y} = \frac{G_F^2}{64 \pi^2 E_\nu\, m}\,|M_{\nu e}|^2,  \\ 
\end{equation}
where
\begin{equation*}
y := \frac{T_e}{E_\nu}=\frac{m}{E_{\nu}}\frac{2 \cos^{2}{\theta_e}}{(1+\frac{m}{E_{\nu}})^{2}-\cos^{2}{\theta_{e}}} 
\end{equation*}
is the ratio of the kinetic energy of the recoil electron $T_{e}$  to the incoming $\nu_e$  energy $E_{\nu}$, and $m$ is the electron mass.
The squared amplitudes have the form  
\begin{eqnarray}
|M_{\rm VA}|^2 &=& \frac{-16 m^2 }{m+y E_{\nu }} \sqrt{\frac{E_{\nu }}{2 m+y E_{\nu }}} \left\{-\left(\sqrt{y} \left(\sqrt{y E_{\nu } \left(2 m+y E_{\nu }\right)} \left(\hat{s}_1\cdot \vec{s}+1\right) \right.\right.\right.\nonumber\\
&& \left.\left.\left. +2 (y-2)\, \, \hat{p}\cdot \vec{s} \,\,E_{\nu }\right) m^2+E_{\nu } \left((y-2) \, \hat{p}\cdot \vec{s} \,\, E_{\nu } y^{3/2}+\left(2 \left(y^2-y+1\right) + \right.\right.\right.\right. \nonumber \\
&& \left.\left.\left.\left. 
\left(y^2+2 y-2\right) \hat{s}_1\cdot \vec{s}\right)
\sqrt{E_{\nu } \left(2 m+y E_{\nu }\right)}\right) m+y \left((y-2) y \right. \right.\right. \nonumber \\
&&\left.\left. \left.
+ 2 (y-1) \hat{s}_1\cdot \vec{s}+2\right)\sqrt{E_{\nu }^5 \left(2 m+y E_{\nu }\right)}\right) c_{\rm A}^2 
+2 c_{\rm V} c_{\rm A}E_{\nu } \left(2 \sqrt{y} \, \hat{p}\cdot \vec{s} \, m^2 + \right.\right. \nonumber \\
&& \left.\left.
\left((y-2) \left(y+\hat{p}_{\nu }\cdot \vec{s}\right) \sqrt{E_{\nu } \left(2 m+y E_{\nu }\right)} 
-(y-4) \sqrt{y} \, \hat{p}\cdot \vec{s} \,\, E_{\nu }\right) m \right. \right. \nonumber \\
&& \left.\left. 
+(y-2) y \left(\left(y+\hat{p}_{\nu }\cdot \vec{s}\right) \sqrt{E_{\nu }^3 \left(2 m+y E_{\nu }\right)}-\sqrt{y} \, \hat{p}\cdot \vec{s} \,\, E_{\nu }^2\right)\right) \right. \nonumber \\
&& \left. + c_{\rm V}^2 \biggl(y \left(\sqrt{E_{\nu } \left(2 m+y E_{\nu }\right)} \left(\hat{s}_1\cdot \vec{s}+1\right)
+2 \sqrt{y} \, \hat{p}\cdot \vec{s} \,\, E_{\nu }\right) m^2 \right. \nonumber \\
&& \left.\left. 
+E_{\nu } \left((y+4) \, \hat{p}\cdot \vec{s} \,\, E_{\nu } y^{3/2}
+\left(-2 \hat{p}_{\nu }\cdot \vec{s} y+2 y+(y (y+2)-2) \hat{s}_1\cdot \vec{s}-2\right) \right. \right.\right. \nonumber \\
&& \left.\left.\left.
\sqrt{E_{\nu } \left(2 m+y E_{\nu }\right)}\right) m - y \Bigl(\left(y \left(y+2 \hat{p}_{\nu }\cdot \vec{s}-2\right)-2 (y-1) 
\hat{s}_1\cdot \vec{s}+2\right) \right.\right.\nonumber \\
&&\left.
\sqrt{E_{\nu }^5 \left(2 m + y E_{\nu }\right)}
 -2 y^{3/2} \, \hat{p}\cdot \vec{s} \,\, E_{\nu }^3\Bigr)\biggr)\right\}
\end{eqnarray}
\begin{eqnarray}
|M_{\rm S_{\rm L}}|^2 &=& 8 m^2 y \left|c_{S_R}\right|{}^2 
\left(\hat{s}_1\cdot \vec{s}+1\right) E_{\nu } \left(2 m+y E_{\nu }\right)
\end{eqnarray}
\begin{eqnarray}
|M_{\rm T_{\rm L}}|^2 &=& \frac{-16 m \left| c_{T_R}\right|{}^2}{m+y E_{\nu }}
\sqrt{\frac{E_{\nu}}{2 m+y E_{\nu}}} 
\biggl(-(y-2) y \left(\left(y+2 \hat{p}_{\nu}\cdot \vec{s}-2\right) \sqrt{2 m+y E_{\nu }} \right. \nonumber \\
&& \left.\left.
-2 \, \hat{p}\cdot \vec{s} \, \sqrt{y E_{\nu}}
\right) E_{\nu}^{5/2} +m \left((y-2) (y+4) \sqrt{y E_{\nu}} \, \hat{p}\cdot \vec{s} \,+\left(-2 (y-2) \hat{p}_{\nu}\cdot \vec{s}
 \right.\right.\right. \nonumber \\
&& \left.\left.\left. 
 +y \left(y \hat{s}_1\cdot \vec{s}+4\right)-4\right) \sqrt{2 m+y E_{\nu}}\right) E_{\nu}^{3/2} 
+m^2 \sqrt{y} \left(\sqrt{y E_{\nu } 
\left(2 m+y E_{\nu}\right)} \left(\hat{s}_1\cdot \vec{s}+1\right) \right.\right. \nonumber \\
&&
+2 (y-2) \, \hat{p}\cdot \vec{s} \, E_{\nu}\Bigr)\biggr)
\end{eqnarray}
\begin{eqnarray}
2\Re(\rm M_{S_L} M^*_{T_L}) &=& \frac{-16 m \sqrt{y} E_{\nu }^{3/2}}{m+y E_{\nu }} \biggl( m \left(\Im\left(c_{S_R}\right) \Im\left(c_{T_R}\right)+\Re\left(c_{S_R}\right) \Re\left(c_{T_R}\right)\right)
\Bigl(-\left(\hat{s}_1\cdot \vec{s}+1\right) \nonumber \\
&& \left. E_{\nu }^{3/2} y^{5/2} + 2 \left(\hat{s}_1\cdot \vec{s}+1\right) \left(y E_{\nu }\right){}^{3/2}
-m \left((y-2) \sqrt{2 m+y E_{\nu }} \, \hat{p}\cdot \vec{s} \right.\right. \nonumber \\
&&\left.\left. 
+\hat{s}_1\cdot \vec{s} \left(\sqrt{y^3 E_{\nu }}-2 \sqrt{y E_{\nu }}
\right) +\sqrt{y^3 E_{\nu }}\right)+2 m \sqrt{y E_{\nu }}\right)
-\left(\Im\left(c_{T_R}\right) \Re\left(c_{S_R}\right) \right. \nonumber \\
&& \left.
-\Im\left(c_{S_R}\right) \Re\left(c_{T_R}\right)\right) \left(m+y E_{\nu }\right) 
\left(\, \hat{p}\cdot \vec{s} \, \hat{p}_{\nu }\cdot (\hat{p}\times \hat{s}_1) \sqrt{y E_{\nu }} \left(2 m+y E_{\nu }\right) \right. \nonumber \\
&&\left.
+\hat{p}_{\nu }\cdot (\hat{s}_1\times \vec{s}) \sqrt{y E_{\nu }} \left(2 m+y E_{\nu }\right) + \left(2 m \left(\hat{p}_{\nu }\cdot (\hat{p}\times \vec{s})+
\hat{p}_{\nu }\cdot (\hat{p}\times \hat{s}_1)\right) \right.\right.\nonumber \\
&& \left.\left. -\hat{p}\cdot (\hat{s}_1\times \vec{s}) \left(2 m+y E_{\nu }\right)\right) 
\sqrt{2 m+y E_{\nu }}\right)\biggr)
\end{eqnarray}
\beq
|M_{\rm T_{\rm R}}|^2 &=& 0, \quad  |M_{\rm S_{\rm R}}|^2 = 0 \nonumber \\
2\Re(\rm M_{S_R} M^*_{T_R}) &=& 0,\quad 2\Re(\rm M_{S_L} M^*_{S_R}) = 0 \nonumber \\
2\Re(\rm M_{S_R} M^*_{T_L}) &=& 0,\quad 2\Re(\rm M_{T_L} M^*_{T_R}) = 0 \nonumber \\
2\Re(\rm M_{S_L} M^*_{T_R}) &=& 0,\quad 2\Re(\rm M_{VA} M^*_{T_L}) = 0 \nonumber \\
2\Re(\rm M_{VA} M^*_{S_L}) &=& 0,\quad 2\Re(\rm M_{VA} M^*_{T_R}) = 0 \nonumber \\
2\Re(\rm M_{VA} M^*_{S_R}) &=& 0
\eeq

If we take the difference of the cross sections $\Omega=\frac12\left[\frac{d^2 \sigma}{d\varphi d \,y}|_{\hat{s}_1}-\frac{d^2 \sigma}{d\varphi d \,y}|_{(-\hat{s}_1)}\right]$  for two opposite directions of target polarization ($\pm \hat{s}_1$) we obtain  
\begin{eqnarray}
\label{a1VA}|M_{\rm VA}|^2 &=&
16 m^2  \hat{s}_1\cdot \vec{s} \left(c_A^2-c_V^2\right) 
 E_{\nu } \left(m y+2 (y-1) E_{\nu}\right)
\\
|M_{\rm S_L}|^2 &=&
8 m^2 y \left| c_{S_R}\right|^2  \hat{s}_1\cdot \vec{s}\, E_{\nu } 
\left(2 m+y E_{\nu}\right)
\\
|M_{\rm T_L}|^2 &=&
-16 m^3 y \left|c_{T_R}\right|^2  \hat{s}_1\cdot \vec{s} \,E_{\nu }
\\
\nonumber\\
2\Re(\rm M_{S_L} M^*_{T_L}) &=& \frac{-16 m \sqrt{y}  E_{\nu }^{3/2}}{m+y E_{\nu }}
\Bigl( m \, \hat{s}_ 1\cdot \vec{s} \,
\left(\Im\left(c_{S_R}\right) \Im\left(c_{T_R}\right)+\Re\left(c_{S_R}\right) \Re\left(c_{T_R}\right)\right) 
\nonumber \\
&&  
\left(-E_ {\nu}^{3/2} y^{5/2}+2 \left(y E_{\nu}\right){}^{3/2}+2 m \sqrt{y E_{\nu}}-m \sqrt{y^3 E_{\nu}}\right)
\nonumber \\
&& 
- \left(\Im\left(c_{T_R}\right) \Re\left(c_{S_R}\right)-\Im\left(c_{S_R}\right) \Re\left(c_{T_R}\right)\right)
\left(m+y E_{\nu}\right) \left(2 m+y E_{\nu}\right)  \nonumber \\
&&\label{a1ST} 
\left(\hat{p}_{\nu}\cdot (\hat{s}_1\times \vec{s}) \sqrt{y E_{\nu}}-
\hat{p}\cdot (\hat{s}_1 \times \vec{s}) \sqrt{2 m+y E_{\nu}}\right)\Bigr)
\end{eqnarray}

\section{Scenario  with   $\rm VA, S_L, S_R $ interactions for $\hat{\eta}_{\nu}^{\perp}\neq 0$ }
\label{app2}
The difference of   cross sections 
$\Delta=\frac12\left[\frac{d^2 \sigma}{d\varphi d \,y}|_{\vec{s}}-\frac{d^2 \sigma}{d\varphi d \,y}|_{(-\vec{s})}\right]$ taken for two opposite spin directions of the recoil  electron $\pm\hat{s}$ with $\hat{p}_\nu\cdot \hat{s}_1=0$ when orthonormality  conditions are requested, i.e.   
 $   \hat{p}\cdot \hat{s}_1=0, \hat{p}\cdot \hat{s}=0,  \hat{s}\cdot \hat{s}_1=0$ has the form
\begin{eqnarray}
\label{vs3}
|M_{\rm VA}|^2 &=& -16 c_{\rm V} m^2 E_\nu^2 \left(\hat{p}_{\nu}\cdot \hat{\eta}_{\nu}-1\right)\hat{p}_{\nu}\cdot \vec{s}
\left( c_{\rm V} y - c_{\rm A} \left( y-2 \right)\right) \nonumber \\
|M_{\rm S_R}|^2 &=& 0, \quad |M_{\rm S_L}|^2 = 0, \quad 2\Re(\rm M_{\rm S_L} M^*_{\rm S_R})=0, \quad 2\Re(\rm M_{\rm VA} M^*_{\rm S_L})=0\nonumber \\
2\Re(\rm M_{\rm VA} M^*_{\rm S_R}) &=&
A_0 \, +
A_1 \, \hat{s}_1 \cdot(\hat{\eta}_{\nu}^{\perp}\times \vec{s}) +
A_2 \, \hat{p}_{\nu} \cdot(\hat{\eta}_{\nu}^{\perp}\times \vec{s}) +
A_3 \, \hat{p}\cdot(\hat{\eta}_{\nu}^{\perp}\times \vec{s}) \nonumber \\
&& +A_4 \, \hat{p}_{\nu} \cdot(\hat{s}_1\times \vec{s}) +
A_5 \, \hat{p} \cdot(\hat{s}_1\times \vec{s})
\end{eqnarray}
where the coefficients $A_0, A_1$, $A_2$, $A_3$,  $A_4$,  $A_5$ are given by
\begin{eqnarray}
A_0 &=& -8 m E_{\nu} \Re{(c_{\rm S_R})}\sqrt{y} \Bigl( \vec{s}\cdot \hat{\eta}_{\nu}^{\perp} \, m \sqrt{y}\left( E_{\nu}y
\left(c_{\rm A}-c_{\rm V}\right) + 2 c_{\rm A}m + 2c_{\rm V}E_{\nu} \right)  \nonumber\\
&& \quad 
- c_{\rm A}\, \hat{p}_{\nu}\cdot \vec{s} \, 
\hat{p}\cdot \hat{\eta}_{\nu}^{\perp}\, \sqrt{E_{\nu}\left(2m+E_{\nu}y\right)^3} +
c_{\rm V} \, \hat{p}_{\nu}\cdot \vec{s} \, \hat{p}\cdot \hat{\eta}_{\nu}^{\perp} \, 
y \sqrt{E_{\nu}^3\left(2m+E_{\nu}y\right)} \Bigr)
\nonumber \\
A_1 &=& 8 c_A \Im{(c_{S_R})} m^2 E_{\nu} y \left(2m+E_{\nu} y\right) \nonumber \\
A_2 &=& -8 c_A \Im{(c_{S_R})} m E_{\nu}^2 y \left(2m+E_{\nu} y\right) \nonumber \\
A_3 &=& 8 \Im{(c_{S_R})} m E_{\nu}^{3/2}\sqrt{y \left(2m+E_{\nu} y\right)}\left( 2c_A m +c_A E_{\nu}y + c_V m y \right) \nonumber \\
A_4 &=& 8 c_A \Im{(c_{S_R})} m \hat{p}\cdot\hat{\eta}_{\nu}^{\perp}\sqrt{y}\left(E_{\nu} \left(2m+E_{\nu} y\right)\right)^{3/2} \nonumber \\
A_5 &=& -8 c_A \Im{(c_{S_R})} m E_{\nu}^2\hat{p}\cdot\hat{\eta}_{\nu}^{\perp} y \left(2m+E_{\nu} y\right) 
\end{eqnarray}
If we additionally calculate the differences $\tilde{\Delta}=\frac12\left[\Delta(\hat{s}_1)-\Delta(-\hat{s}_1)\right]$ of the obtained cross sections for two opposite directions of target polarization $\pm\hat{s}_1$, we get 
\begin{eqnarray}
\label{vs3s1}
|M_{\rm VA}|^2 &=& 0 \nonumber \\
2\Re(\rm M_{VA} M^*_{S_R}) &=&
B_1 \, \hat{s}_1 \cdot(\hat{\eta}_{\nu}^{\perp}\times \vec{s}) +
B_2 \, \hat{p}_{\nu} \cdot(\hat{s}_1\times \vec{s}) +
B_3 \, \hat{p} \cdot(\hat{s}_1\times \vec{s})
\end{eqnarray}
where the coefficients are given by $B_1 = A_1$, $B_2 = A_4$, $B_3 = A_5$.




\end{appendices}




\end{document}